# Node Sensing & Dynamic Discovering Routes for Wireless Sensor Networks

[1]Arabinda Nanda
Department of CSE
Krupajal Engineering College
Bhubaneswar, India
aru.nanda@rediffmail.com

[2]Amiya Kumar Rath
Department of CSE & IT
College of Engineering
Bhubaneswar, India
amiyaamiya@rediffmail.com

[3] Saroj Kumar Rout
Department of CSE
Krupajal Engineering College
Bhubaneswar, India
rout-sarojkumar@yahoo.co.in

**Abstract-***The applications of Wireless Sensor Networks (WSN) contain a wide variety of scenarios. In most of them, the network is composed of a significant number of nodes deployed in an extensive area in which not all nodes are directly connected. Then, the data exchange is supported by multihop communications. Routing protocols are in charge of discovering and maintaining the routes in the network. However, the correctness of a particular routing protocol mainly depends on the capabilities of the nodes and on the application requirements. This paper presents a dynamic discover routing method for communication between sensor nodes and a base station in WSN. This method tolerates failures of arbitrary individual nodes in the network (node failure) or a small part of the network (area failure). Each node in the network does only local routing preservation, needs to record only its neighbor nodes' information, and incurs no extra routing overhead during failure free periods. It dynamically discovers new routes when an intermediate node or a small part of the network in the path from a sensor node to a base station fails. In our planned method, every node decides its path based only on local information, such as its parent node and neighbor nodes' routing information. So, it is possible to form a loop in the routing path. We believe that the loop problem in sensor network routing is not as serious as that in the Internet routing or traditional mobile ad-hoc routing. We are trying to find all possible loops and eliminate the loops as far as possible in WSN.*

## 1. INTRODUCTION

A WSN is composed of a large number of tiny autonomous devices, called sensor nodes. A sensor node has limited sensing and computational capabilities and can communicate only in short distances. Routing protocol is a set of rules defining the way router machines find the way that packets containing information have to follow to reach the anticipated destination.

The concept of WSN is based on a simple equation:

Sensing + CPU + Radio = Thousands of potential applications

As soon as people understand the capabilities of a WSN, hundreds of applications come to mind. Actually combining sensors, radios, and CPU's into an effective WSN requires a detailed understanding of the both capabilities and limitations of each of the essential hardware components, as well as a detailed understanding of modern networking technologies and distributed systems theory's that combines data sensing, computing, and communication has been gaining great popularity in recent years. Several real world applications have already been designed, implemented and deployed [1]. WSN consists of a large number of Sensor Nodes and one or more Base Stations. A Base Station acts as a gateway to connect a WSN to the outside world. Individual Sensor Nodes sense their environment, and transmit the sensed data to a Base Station through a multi-hop network consisting of several sensor nodes. The Base Station in turn transfers the data to the WSN users. Several routing protocol for WSN has been proposed [2, 3, 4, 5, 6, 7, 8, 9, 10].

Deng, Han&Mishra had done a lot of work on routing mechanism for WSN.They had studied on loops and how to eliminate the loops in WSN.But here we add loop finding algorithm, to find all possible loops in WSN. Although each individual sensor node is highly constrained in its computing and communication capabilities, a complete WSN is capable of performing complex tasks.

Common failures in the system includes: *Node Failure, Area Failure and Lost Message.* In order to function properly, the rest of the system must (1) detect failures; (2) determine the cause, such as identifying the types of failure and the failed component; (3) reconfigure the system so that it can continue to operate; and (4) recover when the failed component is repaired.

A node engaged in a handshaking protocol of some kind usually experiences a failure as the lack of the expected response from its partner within a prescribed time limit. The use of time-outs is a common technique for detecting missing response. However, the choice of a specific time-out value presents some practical problems. Too long a time-out result in slow detection of missing message. On the other hand, too short a time-out may trigger false alarms by declaring as missing message that is just delayed. Moreover, short time-outs require the communication subsystem to deal with duplicate message sent in response to hurriedly requested replays.

Every node in a WSN has similar chances to suffer from an arbitrary Node Failure, which is generally caused by battery drain or some internal problem in the node. An Area Failure results in a failure of all nodes within a certain geographical area. This is typically caused by outside accidents, such as a bomb blast, fire, successful denial-of-service attacks, and so on. Figure 1 illustrates these two types of failures in a WSN.

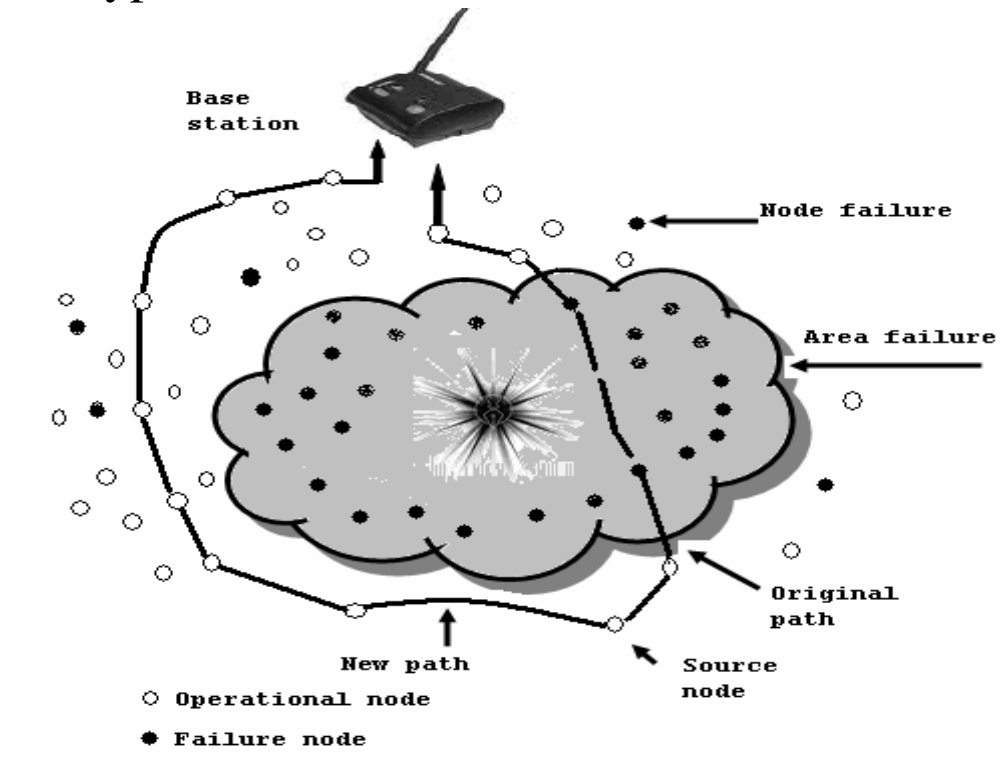

**Figure 1 Node failure in WSN**

Three different methods have been used to maintain routing paths in the occurrence of node failures. In the first method, routing paths are reconstructed from time to time. For example, in a simple beacon protocol [11], a base station periodically broadcasts a beacon message. By receiving a beacon message, a node receives an up-to-date routing path to the base station. Reconstruction of routing paths is expensive in this method and consumes lots of energy. In addition, since reconstruction is not on demand, a node has to wait until the beacon to update the routing information on a node failure.

In the second method, multiple routing paths are used to transfer data. The idea is that unless every path from a sensor node to a base station is broken by a failed node, data can be transmitted to base station. The multipath version of directed diffusion [12] uses this strategy. This method can result in increased energy consumption and packet collisions, because data is sent along multiple paths, irrespective of whether there is a node failure or not. Also, this method cannot guarantee bypassing an area failure.

In the third method a routing path is selected probabilistically. In this method, a node chooses another node to forward a packet with certain probability. Since there is no fixed path to forward data, a failed node can't block all packets from a sensor node to a base station. The ARRIVE routing protocol [13] uses this strategy to forward multiple copies of the same data.

Woo, Tong, and cullar [14] investigated the challenges of multihop routing in wireless sensor networks and proposed a routing scheme based on node's neighborhood link estimates. This protocol is for a many-to-one, data collection routing development in WSN. A sensor network can quickly respond to node failures and data transmission range changes, and find new routing path for sensor nodes. To do this, a sensor node needs to periodically broadcast its routing information, or periodically search its neighbor nodes' routing information. In addition, it needs to maintain a table which contains its neighbor nodes' routing information.

In this paper, we propose a dynamic discovering routing method that can be integrated in any routing protocol for WSN to make it fault tolerant. It dynamically repairs a routing path between a sensor node and a base station. In contrast to [14], a node stores only its parent node routing information, and asks for neighbor nodes routing information when parent node is hard to find. When an original routing path is broken, a node selects a new path from its neighbor nodes. This dynamic discovering routing method tolerates both arbitrary node and area failures.

## 2. PROTOCOL EXPLANATION

### 2.1. Assumptions

In this paper, we center of attention on how each sensor node maintains its routing path to base stations. We assume that the initial routing method from each sensor node to a base station has already been set up. This can be done using a number of protocols that have been proposed in the past, e.g. the TinyOS beacon protocol discussed below. In particular, we assume that each node already has a path to the base station, and knows its parent node, neighbor nodes and the number of hops it is from the base station. This information can be initialized by using the TinyOS beacon protocol for setting up routing paths. In this protocol, the base station floods a beacon message in the network. When a node first knows the beacon message, it records the sender of that beacon message as its parent node and forwards the beacon message to all of its neighbor nodes. When a node needs to send/forward a message to the base station, it sends the message to its parent node. The parent node in turn forwards the message to its parent node, and so on, until the message gets to base station. A key problem with this protocol is that it is not error tolerant. If the parent node of a sensor node fails, the sensor node cannot communicate with the base station

## 2.2.Path Repair Algorithm

The basic idea is to repair routing paths in case of arbitrary node or area failures is quite simple: every node monitors its parent node. When it finds that parent node has failed, it asks its neighbor nodes for their connection information. It then chooses a new parent node from its neighbor nodes based on this connection information. As shown in Figure-1, the method can tolerate node failures and routes a message circulating the failed nodes.

This mechanism consists of four parts: the ***failure detection***, ***failure information propagation***, ***new parent detection***, **and** ***new parent selection***. First, a node detects if its parent node is alive and if the parent node can connect to base station. This part is called *failure detection*. If a node s detects that its parent node works well, it won't do any maintenance work. If there are some problems in parent node, such as node failure or disconnected to base station (possibly one of parent node's ancestor node is failed), node s informs its children nodes about the failure, which is called *failure information propagation*. In addition, s requests the connection information from its neighbor nodes since it needs to choose a new parent node from them. This part is called *new parent detection*. After collecting information from its neighbor nodes, s decides a new parent node based on the information it collected. This part is called *new parent selection*.

We denote **a** as the node who tries to maintain its route path. Node **p(a)** is **a**'s parent node.

1. Node **a** sends FORWARD message to its parent node **p(a)** , and set a timeout (timeout_ppt) for BACK message from **p(a)**.

FORWARD: **a** → **p(a)** :forward_ppt

**2.** If p **(a)** receives the FORWARD message, it will reply a BACK message. The BACK message contains the information that whether **p(a)** connects to base station or not, and if it is connected, the hops to base station. If p (a) connects to base station, it sends BACK_Y message back to **a**.

BACK_ Y: p (a) → a: connect||hops

If **p (a)** cannot connect to base station, it sends BACK_N message back to **a**:

BACK_N: **p (a)** → **a**: broken||broken_ hops

If p (a) cannot connect to its parent node p.parent, the p.broken_hops is set to 1. Otherwise,

P.broken_hops= p.parent.broken_hops + 1

**3 (a).** If **a** receives BACK_Y from **p(a)**, **a** resets its hops as parent p's hops plus one:
**a**hops ← hops + 1. If **a** node's hops beyond a maximum threshold value, it sets itself unconnected: **a**hops ← ∞.

**(b).** If **p (a)** is dead or its signal is blocked, it cannot reply BACK message within timeout. If **a** cannot receive BACK message from **p(a)** within the specified timeout, **a** knows that it cannot connect to base station through **p(a)**. Then it broadcasts a REQUEST message to all of its neighbor nodes to find a new parent node.

REQUEST: **a**→NEIGHBOR: request_parent

**(c).** If **a** receives BACK_N from **p(a)** , **a** knows that **p(a)** cannot connect to base station at that moment. Instead of broadcasting REQUEST message immediately, **a** waits a timeout before sending REQUEST. The timeout depends on the value of broken_hops from BACK_N message. This strategy gives parent node **p** some time to find its new parent node. **a** will set its broken_hop, and propagate it when its children nodes send FORWARD message to **a**.

**4.** when one of **a**'s neighbor node **n** receives REQUEST message from **a**, and if **n** can connect to base station, it sends a REPLY message back to **a**. REPLY message contains the ID of **n**'s parent node, and **n**'s hops to base station:

REPLY: **n** → **a**: connect||**n**_hops||n.parent

If **n** can't connect to base station, it will not send any message back to **a**. Instead, it records **a** as one of its REQUEST senders. (Here, **a**'s children nodes will not sends REPLY message back to **a** since it is not necessary.)
If **a** has not got any REPLY message from its neighbor nodes, it will resend REQUEST after a certain timeout.

**5.** When **a** receives REPLY messages from its neighbor nodes, if the REPLY message says that the sender connects to base station, **a** records the sender as a parent candidate. Finally, **a** selects its new parent node whose hops to base station is smallest among all candidates. After it selects parent node, **a** sets its hops as its parent node's hops plus one:

$$\mathbf{a}_{hops} \leftarrow \mathbf{p(a)}_{hops} + 1.$$

If **a** ever received REQUEST message from its neighbor nodes, it will send REPLY back to the REQUEST senders.

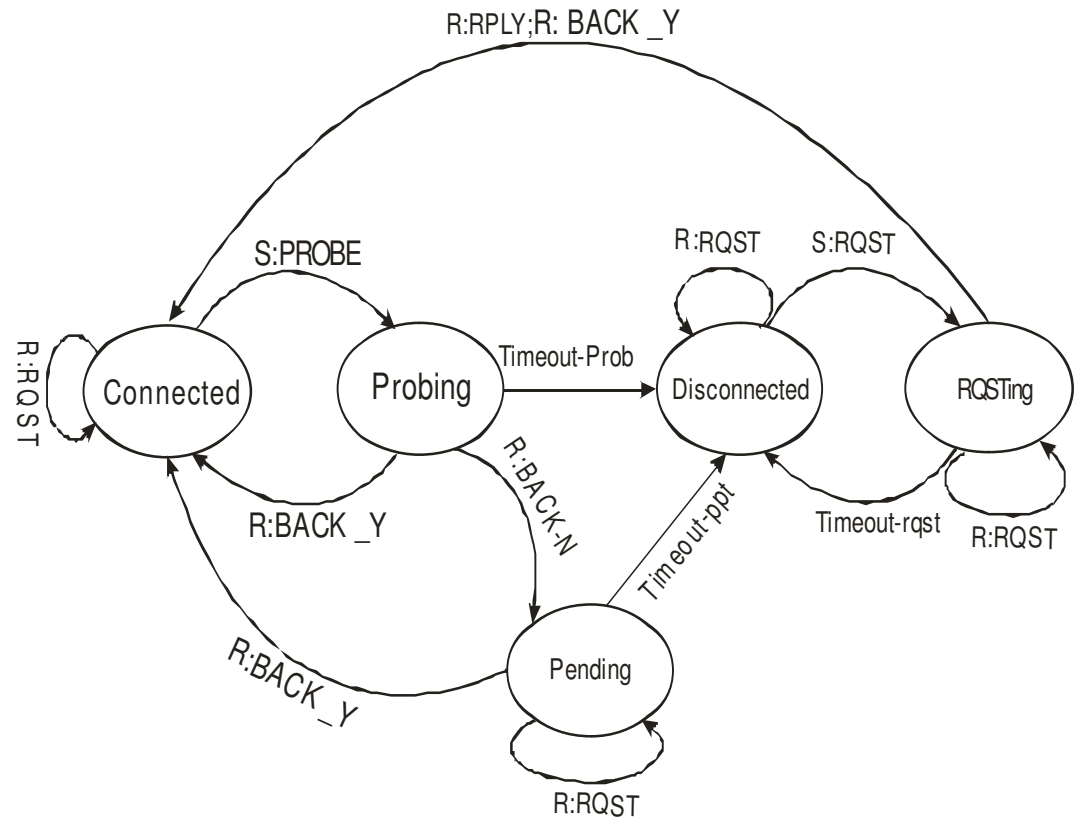

**Figure 2 Protocol Finite Automata**

| | $I_0$ | $I_1$ | $I_2$ | $I_3$ | $I_4$ | $I_5$ | $I_6$ | $I_7$ | $I_8$ |
|---|---|---|---|---|---|---|---|---|---|
| $Q_0$ | $Q_0$ | | | | $Q_1$ | | | | |
| $Q_1$ | | $Q_0$ | $Q_3$ | | | | | $Q_2$ | |
| $Q_2$ | $Q_2$ | | | | | $Q_4$ | | | |
| $Q_3$ | $Q_3$ | $Q_0$ | | | | | $Q_2$ | | |
| $Q_4$ | $Q_4$ | $Q_0$ | | $Q_0$ | | | | | $Q_2$ |

**(Transition table for FA)**

**The states of FA:**
$Q=\{ Q_0,Q_1,Q_2,Q_3,Q_4 \}$
where
$Q_0$ = Connect, $Q_1$=Probing,$Q_2$=Disconnected,
$Q_3$=Pending,$Q_4$=Requesting
**The input of FA:**
$\sum=\{I_0,I_2,I_3,I_4,I_5,I_6,I_7,I_8\}$
where
$I_0$= R: Request, $I_1$= R: Back_Y,
$I_2$= R: Back_N,
$I_3$= R: Reply, $I_4$= S: Forward, $I_5$= S: Request,
$I_6$= Timeout-ppt, $I_7$ = Timeout-Forward,
$I_8$= Timeout-Request.

In figure 2, we present a formal description of this method with a finite Automata (FA). This FA shows the major state translation except the processing of REQUEST requests. We use x : y to describe the state translation condition. x denotes the action of events denotes the content of message. R means receives a message, S means sends a message

## 3. PROPERTIES

### 3.1. Arbitrary node failure and area failure

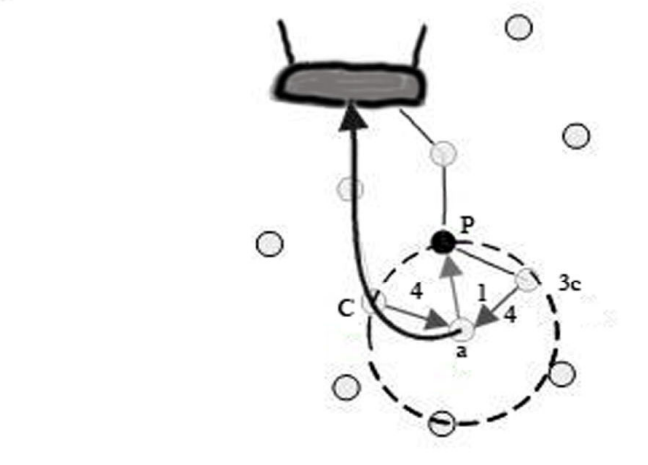

Figure 3. Demo of bypassing a failure node

The proposed method is forceful in finding new paths under arbitrary node failure and area failure. Figure 3 demonstrates how a node find alternative path when its parent node is failed. In figure 3, p(a) is a failed node, showed as a black node. When p(a)'s child node a detects that it cannot connect to p(a) by running step 1, a broadcasts REQUEST message to its neighbor nodes . If any of a's nonchild neighbor nodes can connect their parent nodes, they will send REPLY message back to a. This figure demonstrates that a chooses n as its new parent node from the REPLY messages, and then a has a new path to base station.

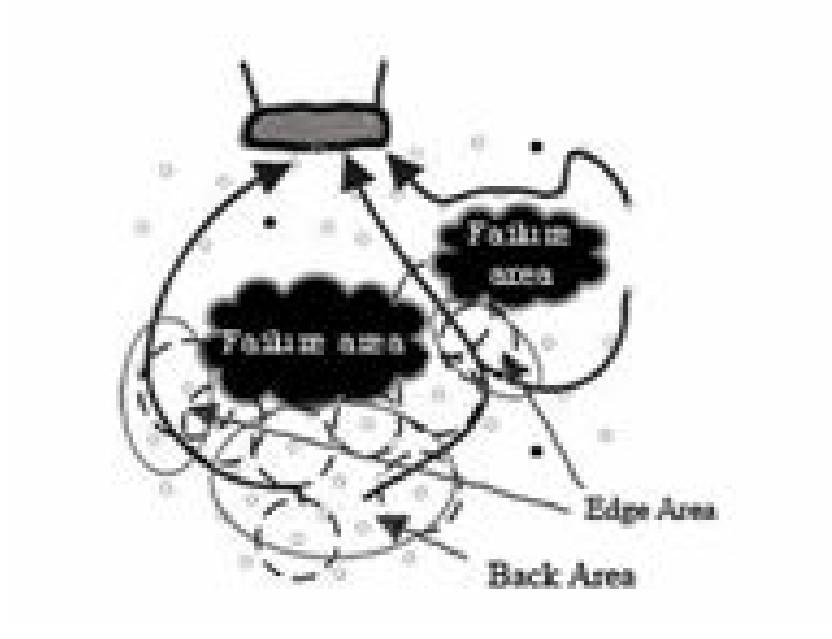

**Figure 4 Bypass Area Failures**

Figure 4 demonstrates that the nodes within a certain area are all failed. This may caused by some accidents, i.e. fire, a bomb, or a signal blocking attack. This type of failure is called area failure. When it happens, the nodes just close to the failure area will send REQUEST messages to their neighbor nodes. In the beginning, some nodes choose other nodes along the failure edge as their parent nodes. That is because these nodes may detect the failure area at slight different time. But quickly, the nodes just behind the failure area will detect that their neighbor nodes are also disconnected to base station. We call this area as “block area”. Because of routing update inconsistency, some nodes may form routing loop in the “block area”. In the edge of the failure area, which we call “edge area”, nodes will find the real path to base station, and the routing information of these nodes will ultimately affect the nodes in “block area” and connect them to base station.

## 3.2. Routing Loop

**3.2.1. Loops:** In our proposed method, every node decides its path based only on local information, such as its parent node and neighbor nodes’ routing information. So, it is possible to form a loop in the routing path, because the REPLY message contains the parent’s node of REPLY sender. A node only finds and eliminates the short loop which is having only 2 or 3 nodes. The longer loops can’t be eliminated. An occurrence of a loop is more likely incase of area failure than arbitrarily node failure. When an area failure occurs, some nodes detect their parent failure and send REQUEST messages, and some nodes that haven’t yet detect failure keep their old routing information. This information inconsistency can create loops. The problem caused by loops is energy consumption and increased packet delay/loss. Nodes in a loop may waste their power by continually forwarding packets.

### 3.2.2 Algorithm for finding all loops in Sensor Network.

An algorithm for finding the loops in a sensor network has been presented. The algorithm first detects a basic loop, copies this as the part of second loop excluding the last element, and then searches in forward and backward directions to find other loops. This process goes on till all the loops are found out. Loop finding is a typical searching process and efficient algorithms for searching loop are not readily available.

**Development of Algorithm**

Input Data: For loop finding, data to be supplied are the basic line information’s, that is, the line with its end nodes. From the raw data, the algorithm will prepare a line-node-incidence matrix (LNI), which will contain the lines connected to a particular node.

The searching process for loop finding will be to start from the source node and go forward till the source node is reached again. To a programmer, however, the problem appears to be a little bit harder because at every step of the searching process, the searching direction has to be chosen judiciously with a wrong direction, the program may enter ending searching process and will never reach the source node, may not be able to find out all the loops or may travel along the same loop every time.

The program must therefore remember the part along which it had already traveled. A line on the other hand, may participate in several loops. The program thus has to select the lines through which it must travel again, though already traveled while finding out other loops. The problem may become easier to understand with the help of an example. The developed algorithm has been tested with fairly large size sensor networks; the network of fig 5 has been taken as example for simplicity. The network of Figure 5 has 5 nodes, eight lines and 10 loops starting from node 1.

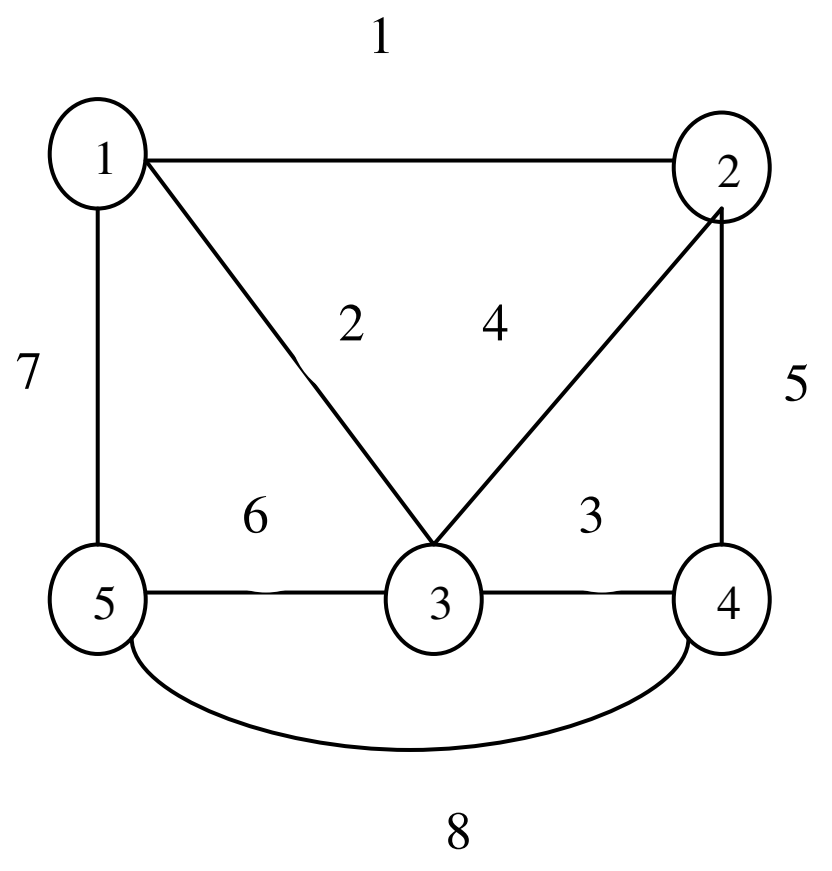

**Figure 5 Example Network**

A list of the loops is given as below:

1→2→3→1
1→2→3→4→5→1
1→2→3→5→1
1→2→4→3→1
1→2→4→3→5→1
1→2→4→5→1
1→2→4→5→3→1
1→3→4→5→1
1→3→2→4→5→1
1→3→5→1

Let node 1 is the source node. Starting from node 1, one may reach node 2 and come back to the source node along line 2 or 7. There are seven such loops with a common second node. All such loops with a common second node will be referred to as a BLOCK. Once all the loops of the first BLOCK are found out, the starting line (Line 1, here) must be omitted because all the possible loops with this line have been found out. From source node one may now reach at node 3 and come back to node 1 through line 7. There are three such loops, forming another BLOCK. Now line 2 will also be omitted and no other loop is possible with only one line connected to source node.

For a network having N lines connected to the source node there will be N – 1 BLOCKS. While the program is in a particular BLOCK, it must store the number of lines, along which it had to travel to complete each loop. All such lines are stored in LECON matrix. But a line may be the part of many loops. As line 4 (3 – 2) is appearing in 3 loops, line 7 (5 – 1) is appearing in 4 loops of the first BLOCK. The program thus defines another matrix LCON, which initially contains the same content as LECON, but afterwards, on reaching a particular node it judiciously select some lines connected to that node to make free and these lines are eliminated from LCON of that node.

As one loop is found out, the program copies it in the next row, excluding the last entry. From the last entry of the new row, it then searches for any other path to come back to the source node. If a path is available, a new loop will be formed and again it will be copied. If there is no way to proceed further in the forward direction, the program will move one step backward and search again. If in the process of going backward, the program comes to second column and can't find any forward path, the end of a BLOCK is indicated. Line connecting the first and second column entry of previous BLOCK will never be considered in the next BLOCK. At the start of a new BLOCK LECON is to be initialized again.

A counter NBLOCK counts the number of BLOCKS. When NBLOCK = number of lines connected to the source node, the end of the search process with a source node is indicated.

If all the loop of the network is required, consider next node as the source node and modify LNI matrix to omit the previous source node from the network. At least three nodes are required to form one loop. Thus the loop finding process will be continuing till the reduced network contains only two nodes.

The complete algorithm is given below:

1. Form LNI from the raw data. Set row number $K_1$=1.
2. Set NBLOCK = 1, initialize the loops, set column no $K_2$=1.
3. Enter source node as the first column entry.
4. From the LNI of the source node take the first line, find its end node, modify LNI if both the nodes to omit the first line. Elements of LNI are to be shifted towards left by one position. Enter the end node as an element of the loop.
5. Set LECON = 0 and LCON = 0 for all nodes.
6. Check serially all the lines connected to the second node. If all the lines have been considered go to step 29. If a new line is found, detect its end node.
7. Detect any line connected to the new node. If no line exists go to next step. If the end node of the new line is the start node, end of a loop is indicated. Go to step 9. If the end bus is not the start node, check if the node has already been entered in the loop. If so, consider the next line, otherwise enter the bus in the loop and go on checking till the start node is reached.
8. Go one step backward. Repeat the search from step 7, if column number = 2 go to step 6.
9. Enter the last node in the loop. Enter all the lines in the loop in LECON.
10. Set LEVEL= $K_2$ – 1.
11. $K_1$ = $K_1$ + 1. Copy the last loop excluding the last element.
12. Set LCON = LECON, NRESTNODE = last element of the present row. $K_2$ = K2 – 1.
13. If number of rows in the present BLOCK is less than 3, go to step 22, otherwise go to next step.

14. $LBS_1$=Loop ($K_{1-1}$, K2), $LBS_2$= Loop ($K_{1-2}$, $K_2$). If $LBS_1$ = $LBS_2$ go to step 22.
15. Set ITN= 0.
16. If $LBS_2$=Loop ($K_{1-1}$, $K_{2+1}$), go to step 19, otherwise go to next step.
17. If $LBS_2$=any element of the present loop go to step 19, otherwise go to next step.
18. Find the line connecting $LBS_1$ and $LBS_2$ if any, if no line exists go to step 19. If any line exists make this line free (eliminate from LCON).
19. If ITN = 1, go to step 22, otherwise set ITN = 1.
20. Check the next column in the previous row. If it is the source node go to step 22.
21. Make $LBS_2$=Source node. Go to step 18.
22. Click if the present loop up to the present entry is same as any other previous loop. If same, go to step 23. Otherwise, make LCON of present node = 0.
23. Take a line from LNI of NRESTNODE. Check if it is in LCON. If no line connected to the node is free go to step 27, otherwise go to next step.
24. Find the end node, if end node=any node in the present row or next column in the previous row except the source node go to step 23. Otherwise go to next step.
25. Enter the node in the loop and the line LCON. If the node is not the source node go to step 26. If it is the source node, store the line in the loop in LECON and go to step 11.
26. LEVEL = LEVEL + 1, NRESTNODE = end node; go to step 22.
27. LEVEL = LEVEL – 1; if LEVEL = 2 go to step 6. Otherwise go to next step.
28. Decrease the column number and go to step 12.
29. NBLOCK = NBLOCK + 1; if NBLOCK = number of lines connected to source node – 1, go to next step. Otherwise set $K_2$=1 and go to step 3.
30. Modify LNI to omit the lines connected to the present source node. Check how many nodes are converted to source node. If number of source node = total node – 3, the procedure ends. Otherwise, source node = next node, and go to step 2.

**3.2.3. Loop Elimination:** We don't use sender ID and originating sequence number to detect loop since that requires a node s to memorize lots of history information if there are lots of nodes sending packets to base station through s. In stead, we propose the following mechanism to eliminate loop. Suppose there is a loop $a_1 \rightarrow a_2 \rightarrow \ldots \rightarrow a_k \rightarrow a_1$. This loop exists because there is a node ($a_k$) that finds that its original path is broken and it can connect to $a_1$. Before choosing $a_1$ as its new parent node, $a_k$ needs to broadcast REQUEST to its neighbor nodes, and so $a_{k-1}$ will know $a_k$' s path is broken. If $a_{k-1}$ and its downstream nodes continually inform their downstream nodes the path broken event, the "path broken" information will quickly propagate to all downstream nodes. At the same time, $a_k$ accepts $a_1$ as its new parent node and sends new hops information to its downstream nodes. Although the "new hops" information will eventually propagate to all downstream nodes, its propagation speed is much slower than "path broken" event, since child node gets "new hops" information after it sends FORWARD message and receives BACK _Y. If there is a loop, from $a_1$ through $a_k$ to $a_1$, the "path broken" event will get to $a_1$ and continue to reach $a_k$, and eventually it will catch "new hops" information. At that time, every node on the loop will get "path broken" event and the path of the loop will disappear.

One way to implement above strategy is that in step3c, after receiving BACK_N, node a immediately sends PENDING message to its children nodes. The format of this message is:
PENDING: a→CHILD:pending||pending hops
Initially, a set pending hops as 1. When a node receives PENDING from its parent node, it increases pending hops by 1 and forwards this message to its children nodes immediately. This way, the "path broken" in- formation will spread to all downstream nodes very quickly. Although the PENDING propagation prevents loop, it may also generate the PENDING message storm to downstream nodes.

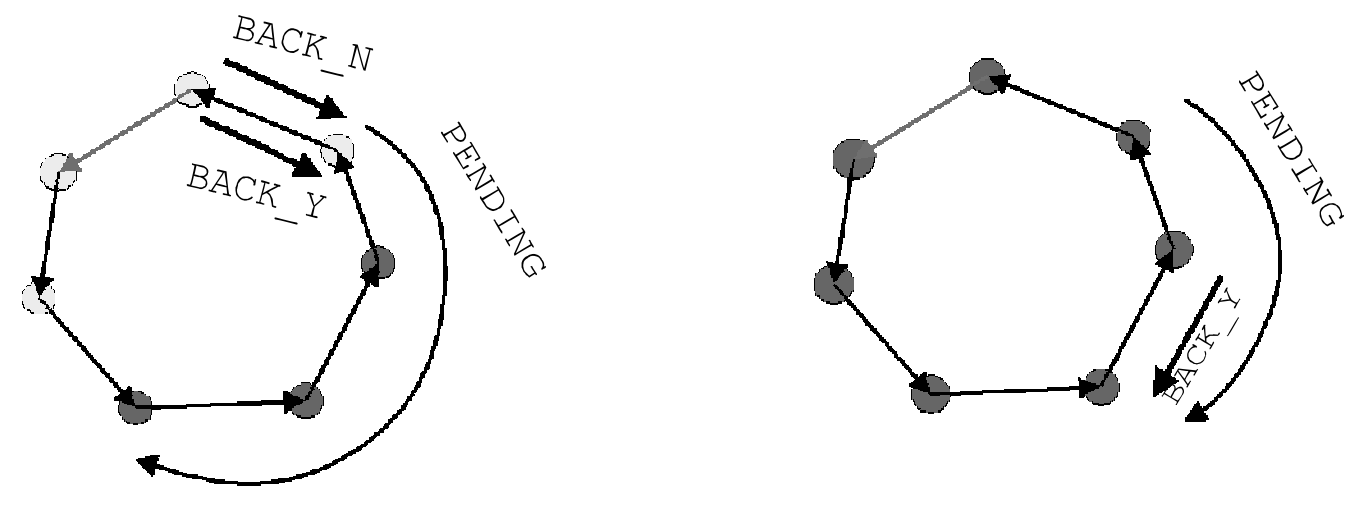

a) Back_N and PENDING is propagated faster than Back_Y

b) PENDING eventually catches up BACK_Y, and every node goes to PENDING state

**Figure 6 Elimination of loop**

## 4. VARIATIONS AND EXTENSIONS

To prevent PENDING storm, a node can slow down PENDING message forwarding. It can wait a short timeout before forwarding a PENDING message. In addition to avoiding loop, a PENDING message can also be used to control packet sending rate, i.e. downstream nodes will slow down or stop sending packets after they know that the path is temporary broken about the failure information. BACK_N and REQUEST messages are used for this purpose. In *new parent detection*, a node finds out information about new parent candidates. REQUEST and REPLY messages are used to find new parent candidates. Finally, in *new parent selection*, a node uses appropriate metrics to choose a new parent node from candidate nodes. In the basic mechanism, a node uses number of hops to base station as the metric.

### 4.1 Simplification of Our Method

In section 2, we described a basic dynamic discovering weight routing algorithm for WSN, which is composed of four parts: *failure detection*, *failure information propagation*, *new parent detection*, and *new parent selection*. In *failure detection*, a node detects if its parent node has failed. We use FORWARD and BACK messages to detect failure. In *failure information propagation*, a node tells other nodes.

### 4.2. Using Diverse Metrics

A central part of our method is the metrics used for new parent node selection. All nodes must use a common metrics to evaluate their routing cost to the base station. This metrics must be such that its value decreases monotonically as you get closer to the base station. Every node can simply use a greedy algorithm to select its parent node based on this metrics value. Number of hops is one type of metric that satisfies this property. Any other metric that satisfies this property can also be used.

#### 4.2.1. Metrics Based on Location Information

If a node can get location of its neighbor nodes (by using GPS, directional antenna or other techniques), then it can choose a parent node based on the location of the failed nodes. For example, when it finds that most of its neighbor nodes in one direction have failed, it concludes that there is an area failure in that direction. In that case, it will choose a new parent node based not only on hops cost, but also on its location relative to the failed area.

### 4.3. Directed Diffusion

Directed Diffusion is an important milestone in the routing research of sensor networks. The idea aims at diffusing data through sensor nodes by using a naming scheme for the data. The main reason behind using such a scheme is to get rid of unnecessary operations of network layer routing in order to save energy. The proposed dynamic discovering routing mechanism can be used in directed diffusion based routing algorithms. In directed diffusion routing algorithm, a destination node disseminates its interest to the network. When corresponding source node gets the interest, it sends events data back to destination along the path through which interest disseminated. Then the destination reinforces a path that connects destination and source nodes. This reinforcement is based on the cached events propagation information. Every new node on the path does reinforcement until the path gets to source node. If a link is broken, a node can find alternate path by running reinforcement again. Since the dissemination of interest passed a large area of nodes between destination and source, the nodes within the area can get and keep the cost metrics to destination. When the reinforced path is broken, other nodes on the path can run this scheme to find another path towards destination.

### 4.4. Node connects and Network reform

In this paper, we focus on node failure problem in WSN. However, we can extend our scheme to deal with new node connects and recovery of failed node. When a new node is added in the network, it broadcasts a message to find its neighbor nodes and their hops to base station. Then this node can choose its parent node and connect the network. In addition, the new node may change other nodes' paths to the base station. Some nodes may have shorter path to the base station through new node. Here, we use conservative strategy for a node to change its parent from a longer path to a shorter path because that is useful to prevent loop, and prevent malicious node from sending fictitious hops information. When a node finds that its neighbor nodes has a shorter hops to base station, it sends two copies of a message to base station, one through that node and the other through its current parent node. If it receives the feedback message from its neighbor node earlier that from its parent node, then it consider to change its parent node.

If there are lots of node failure and new node connecting in the network, our scheme can still build routing paths for alive nodes but the paths may not be efficient. In this situation, it is better to use base station to send beacon message to reconstruct the routing paths in the network.

## 5. CONCLUSION AND FUTURE WORK

In this paper, we have presented a dynamic discovering routing method for communication between sensor nodes and a base station in a WSN. This method tolerates failure of arbitrary individual nodes in the network or a small part of the network by dynamically discovering new routes when nodes fail. The proposed mechanism is generic in the sense that it can be integrated in several routing protocols to make them fault tolerant. In the future, we plan to experiment with this method, including a simulation and implementation, to evaluate its performance and usability in a real sensor network application.

## AUTHORS PROFILE

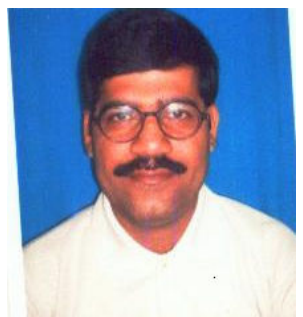

**Prof. Arabinda Nanda:** Received M. Tech (CS) from Utkal University in the year 2007.Currently working as Assistant Professor in the Department of Computer Science & Engineering at Krupajal Engineering College, Bhubaneswar, Orissa, India. Contributed more than 10 research level papers to many National and International journals and conferences. Having research interests include Sensor Network, Adhoc Network, Soft Computing, Artificial Intelligence and Data Mining.

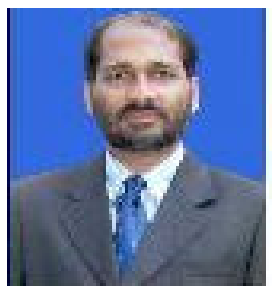

**Prof (Dr) Amiya Kumar Rath:** Obtained B.E. degree in Computer Science & Engg. from Marathwada University, Maharastra in the year 1990, MBA degree in Systems Management from Shivaji University in the year 1993, M.Tech in Computer Science from Utkal University in year 2001 and Ph.D in Computer Science in the year 2005 from Utkal University for the work in the field of Embedded system. Served in various positions in different premier institutes namely College of Engineering, Osmanabad, Maharastra, Orissa Engineering College, Bhubaneswar, Kalinga Institute of Industrial technology (KIIT), Bhubaneswar, Krupajal Engineering College, and Bhubaneswar in the Department CS & IT Engg. Presently working with College of Engineering Bhubaneswar (CEB) as Professor of Computer Science & Engg. Cum Director (A&R) and is actively engaged in conducting Academic, Research and development programs in the field of Computer Science and IT Engg. Contributed more than 30 research level papers to many national and International journals. and conferences Besides this, published 4 books by reputed publishers. Having research interests include Embedded System, Adhoc Network, Sensor Network, Power Minimization, Biclustering, Evolutionary Computation and Data Mining.

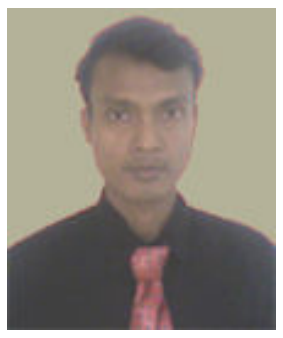

**Prof. Saroj Kumar Rout:** Received M. Tech (CS) from Utkal University in the year 2007.Currently working as Assistant Professor in the Department of Computer Science & Engineering at Krupajal Engineering College, Bhubaneswar, Orissa, India. Contributed more than 08 research level papers to many National and International journals and conferences. Having research interests include Sensor Network, Adhoc Network, Embedded System, and Network Distributed System.